\documentstyle[psfig]{mn}
\newcommand{\kms}{\thinspace\hbox{$\hbox{km}\thinspace\hbox{s}^{-1}$}}
\newcommand{\vsini}{\thinspace\hbox{$v\thinspace\hbox{sin}i$}}
\begin{document}
\title[The mass ratio of EM~Cygni]{A mystery solved: the mass ratio
of the dwarf nova EM~Cygni}

\author[R.C. North, T.R. Marsh, C.K.J. Moran, U. Kolb, R.C. Smith, R. Stehle]
{R.C. North$^1$, T. R. Marsh$^1$, C.K.J. Moran$^1$, U. Kolb$^{2}$,
R.C. Smith$^{3}$ and R. Stehle$^{4}$\\
$^1$University of Southampton, Department of Physics, Highfield,
Southampton SO17 1BJ \\
$^2$The Open University, Department of Physics,Walton Hall, Milton Keynes, MK7 6AA \\
$^3$University of Sussex, Astronomy Centre, Brighton, BN1 9QJ  \\
$^4$University of Leicester, Astronomy Group, University Road, Leicester, LE1 7RH}
\date{Accepted ??
      Received ??
      in original form ??}
\pubyear{1999}

\maketitle

\begin{abstract}

We have discovered that the spectrum of the well-known dwarf nova
EM~Cyg is contaminated by light from a K2--5V star (in addition to the
K-type mass donor star). The K2--5V star contributes approximately 16
per cent of the light from the system and if not taken into account
has a considerable effect upon radial velocity measurements of the
mass donor star. We obtain a new radial velocity amplitude for the
mass donor star of $K_2 = 202\pm3\kms$, which compares with the value
of $K_2 = 135\pm3\kms$ obtained in Stover, Robinson \& Nather's
classic 1981 study of EM~Cyg. The revised value of the amplitude
combined with a measurement of rotational broadening of the mass donor
$\vsini = 140\pm6\kms$, leads to a new mass ratio of $q = M_2/M_1 =
0.88\pm0.05$.  This solves a long standing problem with EM~Cyg because
Stover et al.'s measurements indicated a mass ratio $q > 1$,
a value which should have led to dynamically unstable mass transfer
for the secondary mass deduced by Stover et al. The revised value of
the mass ratio combined with the orbital inclination $i=67\pm2\degr$
leads to masses of 0.99$\pm$0.12M$_{\odot}$ and
1.12$\pm0.08$M$_{\odot}$ for the mass donor and white dwarf
respectively. The mass donor is evolved, since it has a later spectral
type (K3) than its mass would imply.

We discuss whether the K star could be physically associated with EM~Cyg or
not, and present the results of the spectroscopic study.

\end{abstract}


\section{Introduction}
EM~Cygni is one of the most familiar examples of the sub-group of
cataclysmic binary stars called dwarf novae (DN). They consist of a
white dwarf star accreting material through the inner Lagrangian
point from a late spectral-type mass donor. The material spirals down
towards the white dwarf and forms an accretion disc around it. EM~Cyg
belongs to the sub-class of dwarf novae named after the prototype
Z~Cam, in which the peculiar characteristics of the light curve were
first observed. They experience - in addition to the normal
outburst/quiescent states of DN - periods of time when they appear to
be `stuck' in a semi-high state, making them appear visually
approximately $0.7\, {\rm mag}$ brighter than at minimum.

Additionally, EM~Cyg is one of only a handful of cataclysmic variable
(CV) stars which show an eclipse of the disc every orbit and spectral
lines from both stellar components of the binary.  This makes EM~Cyg
special amongst CVs, as it means that accurate binary parameters can
be deduced from spectroscopic and photometric observations.  In a
classic study Stover {\rm et al.} (1981) measured its absorption line
radial velocity amplitude together with that from the emission lines
and obtained a mass ratio $q = M_2/M_1 = 1.26$; this value confirmed
an earlier study of Robinson (1974), who calculated that $q = 1.29$. A
mass ratio greater than one is remarkable because if mass transfer is
from the more massive star to its companion, the orbital separation
must shrink in order to conserve angular momentum. Unless the donor
star shrinks faster than the Roche lobe, mass transfer is
unstable. The instability occurs on a dynamical time scale if the
adiabatic response of the star fails to keep it within its Roche lobe.
At higher donor masses (M$_{2} > 0.8M_{\odot}$) however, the star
shrinks drastically when it loses material at a high rate and so the
mass transfer becomes stable on a dynamical time scale but it is still
unstable thermally (e.g. de Kool, 1992, Politano, 1996). The mass
ratio (q=1.26) and the donor mass (M$_{2}$ = 0.76M$_{\odot}$) of
Stover et al.\ (1981) lead to dynamical instability according to the
criteria given by Politano, 1996: EM~Cyg should not exist.

Despite this unique status, there have been no more recent studies of
EM~Cyg. In part, this is probably because measurements of the radial
velocities of the white dwarfs in cataclysmic variables are normally
based upon the emission lines which usually come from the accretion
disc which surrounds the white dwarf, and these are well known to be
unreliable (Stover et al., 1981). It would perhaps not be surprising if this
problem affected Stover {\rm et al.}'s (1981) study of
EM~Cyg. However, EM~Cyg is one of the longer period dwarf novae
($P=6.98\,{\rm h}$), for which the distortions of the emission lines
are less significant because of the relatively high amplitude of the
white dwarf, and in Stover {\rm et al.}'s study, the emission and
absorption line amplitudes are only $20^\circ$ from being in
anti-phase, by no means a large distortion for these stars.

In this paper we present new spectra of EM~Cyg which show clear
evidence for contamination from a third star, which happens to have a
very similar spectral type to the mass donor star. Accounting for the
presence of this star increases the radial velocity semi-amplitude of
the mass-donor star considerably, and causes the mass ratio ($q =
M_{2}/M_{1}$) to become less than one, at the same time leading to a
larger donor mass. Thus the mass transfer in
EM~Cyg is therefore stable as expected from theory. 

\section{Observations}
\label{sec:obs}
On 22 June 1997 we took 102 spectra with the Intermediate Dispersion
Spectrograph (IDS) on the $2.5\,{\rm m}$ Isaac Newton Telescope (INT)
on the island of La Palma in the Canary Islands. Our dispersion was
0.4 \AA\, per pixel and the spectra covered 6230 to 6650 \AA. The
resulting spectra have a resolution of 0.8 \AA\,(FWHM).  Conditions
were clear throughout the night and the seeing was approximately 1
arcsec.

\begin{figure}
\hspace*{\fill}
\psfig{file=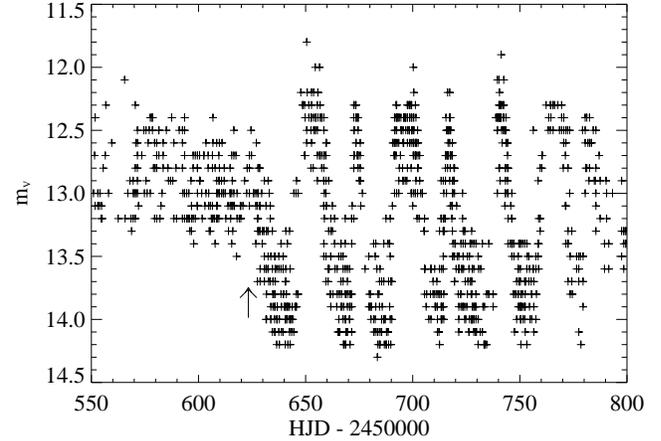,width=85mm}
\hspace*{\fill}
\caption{The light curve of EM~Cyg during 1997. The arrow
marks the date of the observations.}
\label{fig:lig}
\end{figure}

Each target exposure was of $200\,{\rm s}$ duration.  The 0.8 arc~sec
wide spectrograph slit was oriented at a position angle of $255^\circ$
to capture the spectrum of a star 22 arcsec from EM~Cyg in order to
calibrate slit losses. Arc calibration spectra were taken every 35
minutes or so. The arc spectra were fitted with fourth-order
polynomials. The rms scatter of these fits was of the order
0.003\AA. Arc spectra were linearly interpolated in time to provide
the wavelength calibration for EM~Cyg. The CV appeared to be
approaching its low state after a standstill when the spectra were
taken (Mattei, 1999). Fig.~\ref{fig:lig} shows the light curve
behaviour of EM~Cyg during the period April 1997 to February 1998.

The CCD frames were bias-subtracted and then corrected for any
pixel-to-pixel sensitivity variations using exposures of a tungsten
lamp. Variations in the illumination of the slit were corrected using
observations of the twilight sky. We extracted the spectra with
optimised weights (Horne 1986). Spectra of radial-velocity standard K
and M dwarf stars were also obtained. In all, 25 standard stars were
observed, selected from the lists of Marcy et al. (1987), Beavers and
Eitter (1986) and Duquennoy, Mayor and Halbwachs (1991).

\section{Results}
\label{sec:res}

\begin{figure}
\hspace*{\fill}
\psfig{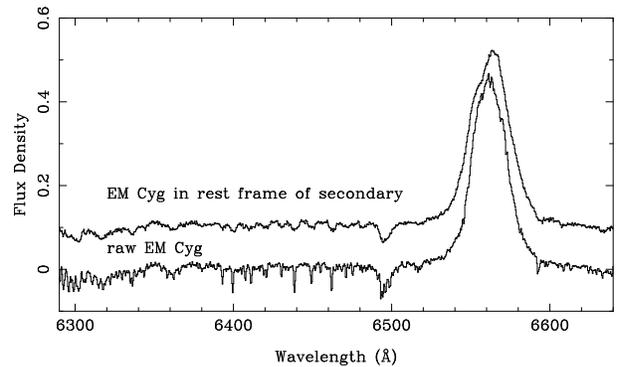}
\hspace*{\fill}
\caption{The mean spectrum of EM~Cyg observed on 22 Jun 1997.}
\label{fig:avspec}
\end{figure}

The mean spectrum of EM~Cyg is presented in Fig.~\ref{fig:avspec}
(lower spectrum). The upper spectrum in Fig.~\ref{fig:avspec} shows
the average of the 102 spectra after subtraction of the contribution
to the absorption lines from the contaminating K-type star (hereafter
`third star'), see section~\ref{sec-anal}, and after each spectrum
has been shifted into the rest frame of the mass donor in EM~Cyg in
order to remove the effect of orbital smearing.  Absorption features
characteristic of a K-type star are visible in the lower
spectrum. They are sharp, which is unexpected because the lines from
the mass donor should be broadened by rotation, and by orbital motion.
These lines are from the third star. It can be seen from this plot
that the absorption lines in the upper spectrum (which represent those
from the mass donor of EM~Cyg) are indeed much wider than those in the
lower spectrum, which is as expected due to the broadening effects
previously noted.

\begin{figure*}
\hspace*{\fill}
\psfig{file=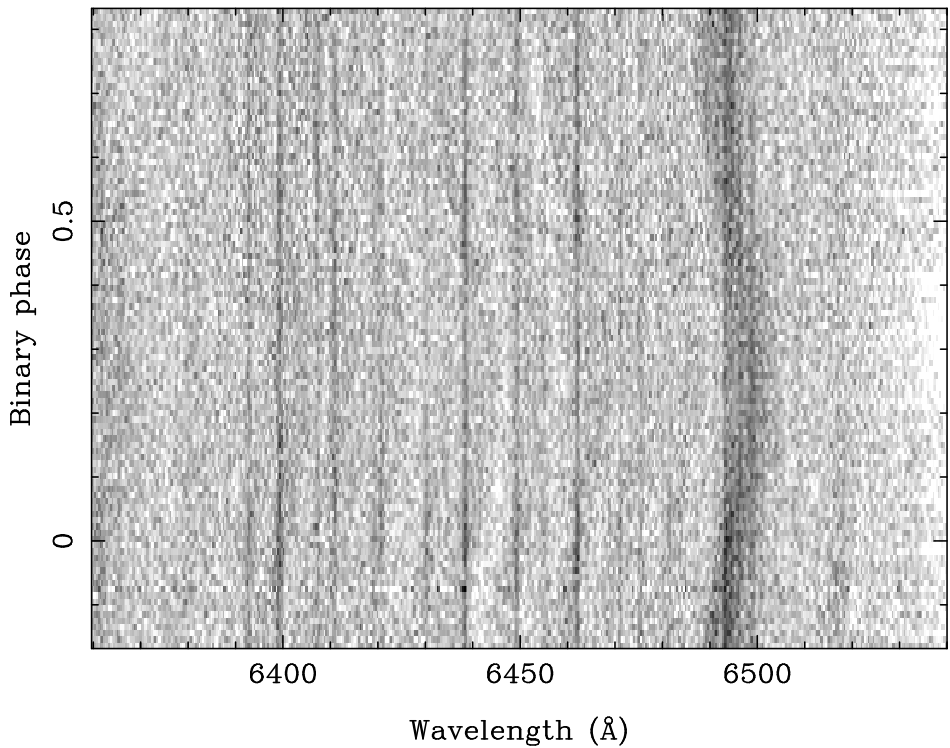,width=125mm}
\hspace*{\fill}
\caption{The trailed spectrum of EM~Cyg scaled to show the absorption 
features. The contaminating K star is visible in narrow lines that run
straight up the plot. The donor star is scarcely visible, but
sinusoidal behaviour can be spotted in the feature at 6495
\AA. H$\alpha$ is not included on this plot (see Fig.~\protect\ref{fig:hatrails}).}
\label{fig:trail}
\end{figure*}

This is confirmed from the trailed spectrum presented in
Fig.~\ref{fig:trail}.

The procedure followed to get from the lower spectrum to the upper
spectrum in Fig.~\ref{fig:avspec} was fairly complex, and so we
start with a summary of the main points.

\begin{enumerate}
\item The continua of the individual EM~Cyg spectra were fitted and
then subtracted. The resulting spectra were then re-binned onto a
uniform velocity scale. The radial-velocity standard stars also had
their continua subtracted, and were then binned onto the same velocity scale. 
\item The average EM~Cyg spectrum was cross-correlated (using the method of
Tonry \& Davis, 1979) with each different 
radial-velocity standard star. This was done in order to obtain a list
of the relative shifts between the two sets of absorption lines. These
shifts were then applied to the standards, so that the absorption
lines were congruous with those of the EM~Cyg spectrum. 
\item The standard star spectra were then scaled and subtracted from
the EM~Cyg spectra in order to remove the absorption lines of the
third star. We refer to this process as ``optimal subtraction''.
\item The spectral type of the standard star which gave the lowest
$\chi^{2}$ on subtraction was adopted as the spectral type of the
third star (e.g. see Fig.~\ref{fig:sptypecon}).
\item The individual, contaminant-free EM~Cyg spectra were then
cross-correlated again with the radial-velocity standards. This time
the standards were artificially broadened to match the rapid rotation
of the mass donor (see section~\ref{sec-rv}).
The resulting  radial velocities were then fitted with 
circular orbits in order to determine the radial velocity
semi-amplitude of the mass donor, and the systemic velocity of
the EM~Cyg.
\item The fit which minimised $\chi^{2}$ was then shifted out of the
individual EM~Cyg spectra in order to correct for the orbital motion
of the mass donor. The resulting spectra were averaged to obtain a
spectrum of EM~Cyg in the rest frame of the mass donor
(Fig.~\ref{fig:avspec}, upper spectrum).
\item In order to derive the best-fit spectral type, and match the
$\vsini$ of the mass donor steps (i) to (iv) were repeated
including the step of artificially broadening the
radial-velocity standards (as in step (v)). See section~\ref{sec-vsini}
for details.
\end{enumerate}

\subsection{Analysis}
\label{sec-anal}

\begin{figure*}
\hspace*{\fill}
\psfig{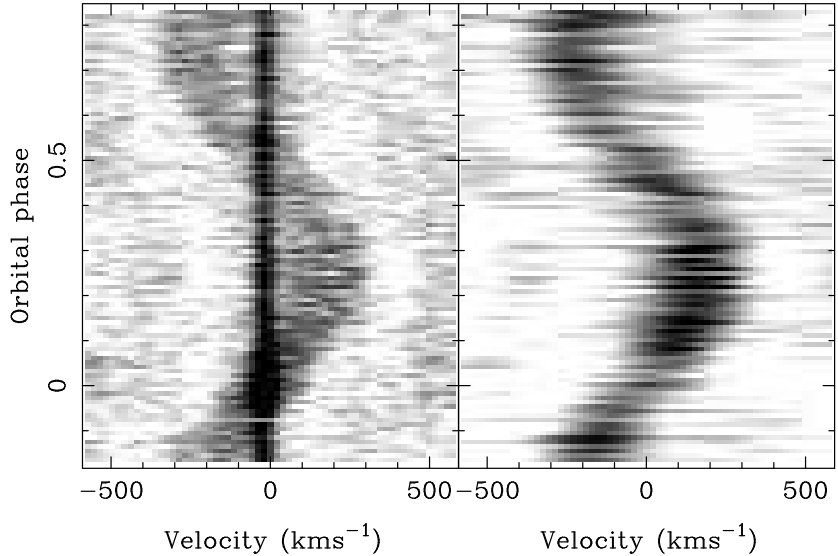}
\hspace*{\fill}
\caption{Trailed spectra of the cross-correlations with a K-type standard
a) before subtraction of a K3V standard, left, and b) after the
subtraction, right. As can be seen from a), the contaminating spectrum
produces another set of correlation peaks, visible here as a straight
line going up the trailed spectrum centred at -20\kms. The correlation
peaks measuring the radial velocities of the mass donor around the
orbit are also visible in both plots. No artificial broadening has
been applied to the standard in plot a), but in plot b) the standard
has been artificially broadened to $\vsini= 140\kms$}
\label{fig:xcor}
\end{figure*}

The presence of the donor star is hard to see in the raw data
(Fig.~\ref{fig:trail}), but becomes clear after cross-correlation with
a K-star radial-velocity standard (Fig.~\ref{fig:xcor}). The cross-correlation
technique used is similar to that used for SS~Cygni (Stover et al.,
1980). The spectra had their continua fitted and then subtracted and
were then transformed onto a logarithmic scale so that the
cross-correlation technique could be used to measure the velocity
shifts due to Doppler effects. In order to effectively remove the
contribution to the absorption lines from the third star, the
relative shifts between the velocity-standard stars and the
average EM~Cyg spectrum were determined, and then applied to the
velocity standards. Then the standards were subtracted from the EM~Cyg
spectra to find the optimum fraction at which the standard should be
scaled; see section~\ref{sec-vsini} for a more detailed description of
the procedures followed.  This routine also provides a formal error on
the fractional contribution from the third star,
and the standard star which gives the minimum $\chi^{2}$ indicates its optimum
spectral type. The standards were then
multiplied by this constant, and subtracted from the EM~Cyg
spectra. The optimum fraction obtained is 0.160$\pm$0.002, with a K3V
standard star.

\begin{figure}
\hspace*{\fill}
\psfig{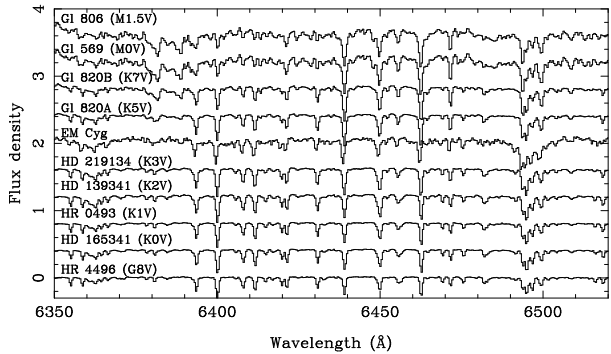}
\hspace*{\fill}
\caption{The average spectrum of EM~Cyg scaled to show the absorption 
lines in the wavelength region of 6350 -- 6540\AA\,, placed in an
apparently suitable position for the third star between spectral types
K3V and K5V. The broadened absorption lines from the mass donor
underlie those from the third star. The sensitivity of the relative
depth of the blends at 6400\AA\, \& 6440\AA\, with effective
temperature is very apparent.}
\label{fig:sptypecon}
\end{figure}

The spectral type of the third star could be initially deduced
from the appearance of the absorption line spectrum between 6350\AA\
and 6540\AA\ (see Fig.~\ref{fig:sptypecon}). The criteria developed by
Casares \& Charles (1993) to determine the spectral type of the mass
donor in V404~Cyg at wavelengths around H$\alpha$ were also used to
classify the contaminant. In particular, the lines at Ca\,{\sevensize
I}\,${\it\lambda\lambda}\/ 6439.1$, and Fe\,{\sevensize
I}\,${\it\lambda\lambda}\/6400.0 + 6400.3$, and the blends at
Ca\,{\sevensize I}$\,{\it\lambda\lambda}\/ 6449.8 + 6450.2$ and
Ca\,{\sevensize I}\,${\it\lambda}\/ 6462.6 +$ Fe\,{\sevensize
I}\,${\it\lambda}\/ 6462.7$, whose relative depths are useful
indicators of spectral type (Casares \& Charles, 1993) were
used. However, as can be seen from Fig.~\ref{fig:sptypecon} the
relative depth of the blends at wavelengths 6400 \AA\, and 6440 \AA\,
are also sensitive to variations in T$_{\rm eff}$\/. The blend at 6400
\AA\, dominates at spectral types around G8V but then diminishes
relative to the blend at 6440 \AA\, becoming equal at around K3, until
it is hardly apparent at spectral types around M4V. In EM~Cygni these
blends appear to have equal strengths relative to each other,
indicating a spectral type around K3V. The optimal subtraction process
described earlier in this section was carried out for each individual
EM~Cyg spectrum, and a different constant computed for each. Then the
standards were scaled using these constants, and subtracted from the
EM~Cyg spectra. The spectral type of the third star was best fitted by
standards between types K2V and K5V. The standard which minimised the
value of $\chi^{2}$ had a spectral type of K3V.

The third star subtraction process was then carried out `in reverse'
to check for possible systematic errors caused by this spectral typing
procedure. This means that the absorption lines from the mass donor in
EM~Cyg (which are broadened by rotational and orbital effects) were
optimally subtracted first (using the same general method as in
section~\ref{sec-rv}), and then the optimum spectral type of the
third star was determined. In this case the optimum spectral type for
the third star came out slightly earlier at K2V, indicating
that we cannot constrain the spectral type any more precisely than
between K2V and K5V with this method.

 Finally the sequence of cross-correlation, shifting, and optimal
subtraction was repeated using the best-fit spectral-type standard to
the contaminating absorption lines (K3V).  The resulting spectra were
then used (see section~\ref{sec-rv}) to find out the spectral type of
the mass donor in EM~Cyg. The relative shift between the standard star
absorption lines and those of EM~Cyg's average spectrum gave a measure
of the systemic velocity of the third star. This turned out to
be $\gamma = -20\pm3\kms$. This is similar to the systemic velocity
derived for the mass donor in EM~Cyg by Stover et al.\ (1981) from the
absorption lines of the spectrum in the H$\beta$ region
($-23\pm6\kms$), and is also similar to the value determined in
section~\ref{sec-rv} ($-25\pm2\kms$). The agreement in systemic
velocities of the mass donor and the third star is a hint that they
are physically associated (see section~\ref{sec-mass}).

The next section details the method used to obtain the radial velocity
semi-amplitude of the mass donor (which incorporates the
cross-correlation technique followed by Stover et al.\ (1980)).

\subsection{Radial velocity of the Donor star}
\label{sec-rv}

\begin{figure}
\hspace*{\fill}
\psfig{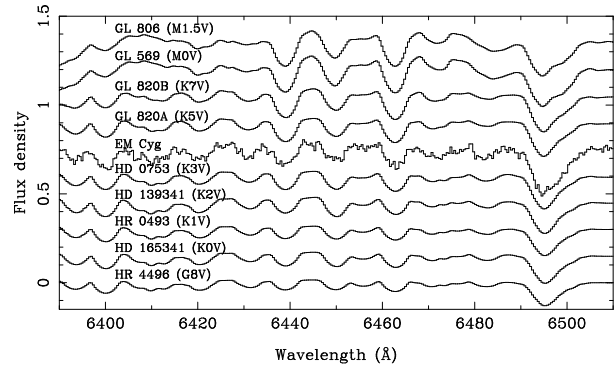}
\hspace*{\fill}
\caption{The average Doppler-corrected spectrum of EM~Cyg after
removal of the third star scaled to show the absorption 
lines in the wavelength region of 6390 -- 6510\AA\,, placed in an
apparently suitable position between spectral types K3V and K5V. The
sensitivity of the relative depth of the blends at 6400\AA\, \& 6440\AA\,
with effective temperature is very apparent.} 
\label{fig:sec}
\end{figure}

Once the individual spectra of EM~Cyg had been corrected for the
presence of the third star, the radial velocity
semi-amplitude of the actual mass donor was calculated.  The corrected
EM~Cyg spectra were cross-correlated (using the method of Tonry \&
Davis, 1979) with the radial-velocity standards each of which
had been artificially broadened. The resulting radial velocity
curves were then fitted with circular orbit fits of the form
\[
V(t) = \gamma + K{\rm sin}\,\frac{2\pi(t-t_{0})}{P} .  \] Then each
EM~Cyg spectrum was shifted to correct for the orbital motion of the
donor star, and the results averaged. This gave the average spectrum
of EM~Cyg in the rest frame of the mass donor (Fig.~\ref{fig:avspec}),
which clearly shows the rotationally broadened absorption lines from
the donor, especially in the wavelength region 6350\AA\,--
6540\AA. Next, this spectrum was cross-correlated once again with the
velocity standards and the relative shifts computed and applied. Then
a constant times each rotationally broadened standard was subtracted
from the average EM~Cyg spectrum. The fraction chosen was that such
that the $\chi^{2}$ of the residual spectrum was minimized. Plotting
the values of $\chi^{2}$ against the spectral type of the standard
star revealed a minimum at spectral types around K3. The circular
orbit fit for the K3 standard was then shifted out of the
contaminant-free EM~Cyg spectra, and the optimal subtraction procedure
then repeated to check for systematic errors which could appear due to
spectral type mismatch. Initially, the spectral type of the mass-donor was
estimated by inspection, in the same manner as for the third
star (see section~\ref{sec-anal}). The absorption line relative
strengths were noted between wavelengths 6350\AA\, and 6540\AA\, as
compared to rotationally broadened versions of the standard
stars. Fig.~\ref{fig:sec} shows the apparent best position to place
EM~Cyg on the plot, in amongst the standard stars whose spectral types
range from G8V to M1.5V. Most of the M dwarfs for which spectra were
obtained have been omitted from this diagram simply because it became
obvious, both by inspection and from the results of the initial
optimal subtraction, that they weren't a suitable spectral-type match
to the mass donor in EM~Cyg. The spectral features around 6400\AA\,
and 6440\AA\, in particular, were noticeably different, as was the
general shape of the spectrum blueward of 6400\AA.

\begin{figure}
\vspace*{-5mm}
\hspace*{\fill}
\psfig{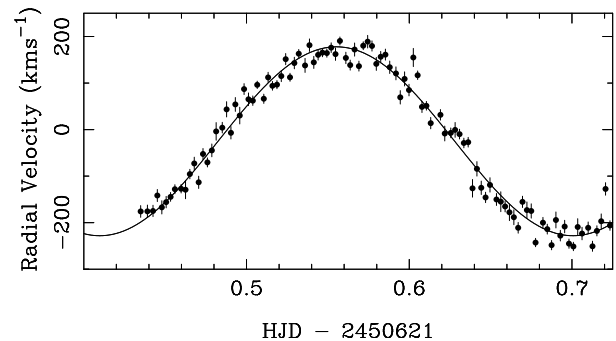}
\hspace*{\fill}
\caption{The 102 measured mass-donor velocities are indicated by the
data points. The solid line is the fit to these points, details of
which are given in Table~\protect\ref{tab:param}.}
\label{fig:fit}
\end{figure}

After this iterative process had been carried out using the standard
with the best fit spectral type (broadened according to the measurement
of the rotational velocity, which is detailed in
section~\ref{sec-vsini}), a revised value for the radial velocity
semi-amplitude of the mass donor star was
obtained. Fig.~\ref{fig:fit} shows the circular orbit fit to all the
data points. The parameters of this fit are given in
Table~\ref{tab:param}. The rms error of the fit is 13$\kms$.

\begin{figure}
\hspace*{\fill}
\psfig{file=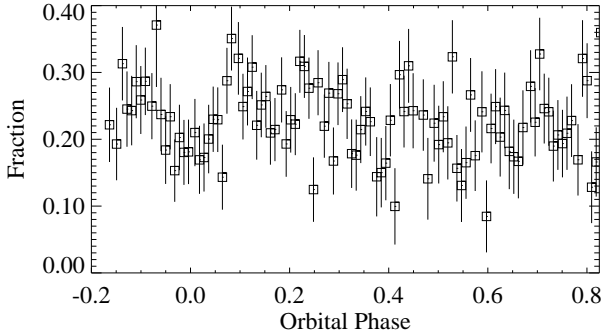,width=90mm}
\hspace*{\fill}
\vspace*{-5mm}
\caption{The fractional contribution to the total light by the donor
star calculated using normalised spectra. For normalised spectra, the
value obtained from the optimal subtraction process for the constant
which the standards are multiplied by is the fractional contribution
of the mass donor star to the total light.}
\label{fig:seccon}
\end{figure}

Fig.~\ref{fig:seccon} shows the fractional contribution from the mass
donor to the total light as a function of orbital phase. There is no
obvious variation in the fraction contributed by the mass donor over
the orbital period. Also, there appears to be no visible decrease
around phase 0.5, which indicates that irradiation of the donor star
by the white dwarf is not important. This is confirmed by the lack of
any obvious distortion of the radial velocity curve (see
Fig.~\ref{fig:fit}), especially around phases 0.25 -- 0.75, compared
with, for example, the distortion seen in the radial velocity curve of
HS1804+6753 (Billington, Marsh \& Dhillon, 1996).  A weighted
least-squares fit to the data points in Fig.~\ref{fig:seccon}
gives the mean fraction contributed by the mass donor over the
orbital period to be 0.231$\pm$0.005.

\begin{table}
\caption{Orbital elements and derived parameters}
\label{tab:param} 
\begin{minipage}{120mm}
\begin{tabular}{lcc}
\hline
Element & Mass Donor\footnote{Reference: this paper} & White Dwarf\footnote{Reference: Stover et
al.\ (1981)} \\
\hline
K\ ($\kms$) & 202$\pm$3 & 170$\pm$10 \\
$\gamma\  (\kms)$ & -25$\pm$2 & -57$\pm$7  \\
$T_{0}\ (JD\odot)$ & 2450621.482$\pm$0.001 &
2443697.128$\pm$0.003  \\
${\rm M\,sin^{3}}\,i\  (M_{\odot})$ & 0.77$\pm$0.08 & 0.48$\pm$0.06 \\
\hline
\end{tabular}
\vspace{-7mm}
\end{minipage}
\end{table}

\subsection{Rotational Velocity of the Donor Star}
\label{sec-vsini}

\begin{figure}
\hspace*{\fill}
\psfig{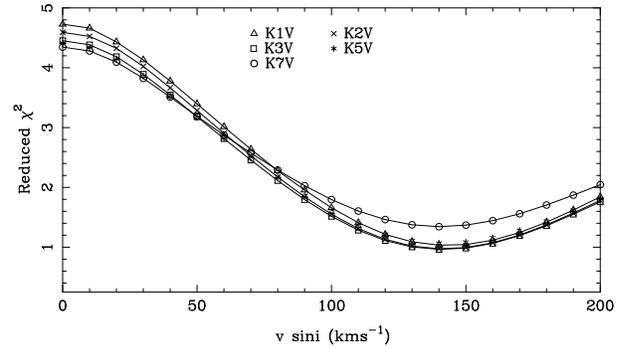}
\hspace*{\fill}
\caption{A plot of the $\chi^{2}$ curves obtained using different
spectral type standards, which show that the minima occur at a
$\vsini$ value around 140$\kms$. }
\label{fig:vsini}
\end{figure}

The rotational broadening of the donor star was estimated by shifting
the radial velocity fit determined in the previous section out of each
EM~Cyg spectrum, and then co-adding the results to produce a spectrum
of the star in the rest frame of the mass donor (see
Fig.~\ref{fig:avspec}). Then this average spectrum was
cross-correlated with the standard stars, whose spectra had been
artificially broadened by a range of velocities (10 -- 200$\kms$ in
steps of 10$\kms$). Both the object and standard star spectra were
normalised in the continuum.  The radial-velocity standard spectra
were then shifted by an amount determined from the correlation. A
constant times each broadened standard spectrum was then subtracted
from the object spectrum, to produce a residual spectrum. Then a
smoothing coefficient was applied to the difference spectrum in order
to remove any large scale features. The $\chi^{2}$ of the result was
computed in the region containing the absorption line features (6390
-- 6510\AA\,). The resulting curves of $\chi^{2}$ versus the value of
the rotational broadening used are shown in Fig.~\ref{fig:vsini}, for
the spectral types from K1 to K7.

From Fig.~\ref{fig:vsini} it is possible to deduce a value for the
rotational broadening and an estimate of the spectral type of the mass
donor. Fig.~\ref{fig:vsini} also shows that the systematic error due
to using different spectral types to obtain values for $\vsini$ is
less than 10$\kms$. The minimum $\chi^{2}$ is obtained for $\vsini= 140\pm3\kms$ and spectral type K3V.

Another systematic effect to be taken account when calculating the
rotational broadening of the mass donor is its non-spherical
shape. The size of this effect was estimated by computing model
absorption profiles including the effects of Roche geometry and system
parameters appropriate for EM~Cyg. These profiles were then used to
broaden a K3 standard star. The $\vsini$'s of the resulting
simulated spectra were then estimated using the same procedure applied
previously to the data. The parameters of Table~\ref{tab:syspar} were
used to define the Roche lobe (i.e. $q=0.88$, $K_{1}+K_{2}=372\kms$),
and the method outlined by Marsh, Robinson \& Wood (1994) was used.
Table ~\ref{tab:ld} shows the results of this process and demonstrates
that the distortion of Roche lobe geometry has a negligible effect since
we recover $\vsini = 140\kms$ for our assumed limb darkening of
0.5. However, one can see that increasing the value of the limb
darkening parameter used to calculate the rotational broadening of the
lines from the mass donor makes the star appear smaller, and
introduces a systematic error of the order 10$\kms$. In all subsequent
analysis the value used for the linear limb-darkening parameter was
0.5.

\begin{table}
\caption{Rotational broadening from model line profiles}
\label{tab:ld} 
\begin{minipage}{120mm}
\begin{tabular}{lcc}
\hline
Inclination  &Limb Darkening\footnote{Gravity darkening
taken to be 0.08 \\ Mass ratio 0.88 assumed} & $\vsini$  \\
 & &  ($\kms$) \\
\hline
69$^{\circ}$  & 0  & 147$\pm$4 \\
69$^{\circ}$  & 0.5 & 140$\pm$3 \\
69$^{\circ}$  & 1 & 130$\pm$3 \\
\hline
\end{tabular}
\vspace{-7mm}
\end{minipage}
\end{table}

In addition, the systematic error introduced by altering the linear
limb-darkening parameter whilst artificially broadening the standard
stars was also accounted for. Using a value of 0 for the linear
limb-darkening parameter with the best fit standard discovered from
Fig.~\ref{fig:vsini}, the measured $\vsini$ dropped to
$135\pm6\kms$. With the limb darkening set to 0.5, the value increased
to $140\pm3\kms$, an error of magnitude 5$\kms$.

Accounting for these possible sources of error, we adopt a value for
the rotational broadening of the mass donor star, $\vsini$, of
$140\pm6\kms$. From the $\vsini$ curves of Fig.~\ref{fig:vsini},
and the relative depths of the absorption lines, the spectral type of
this star is found to be K3V, and it contributes 23 per cent of the total
light in the H$\alpha$ wavelength region.

\subsection{Doppler Tomography}

\begin{figure*}
\hspace*{\fill}
\psfig{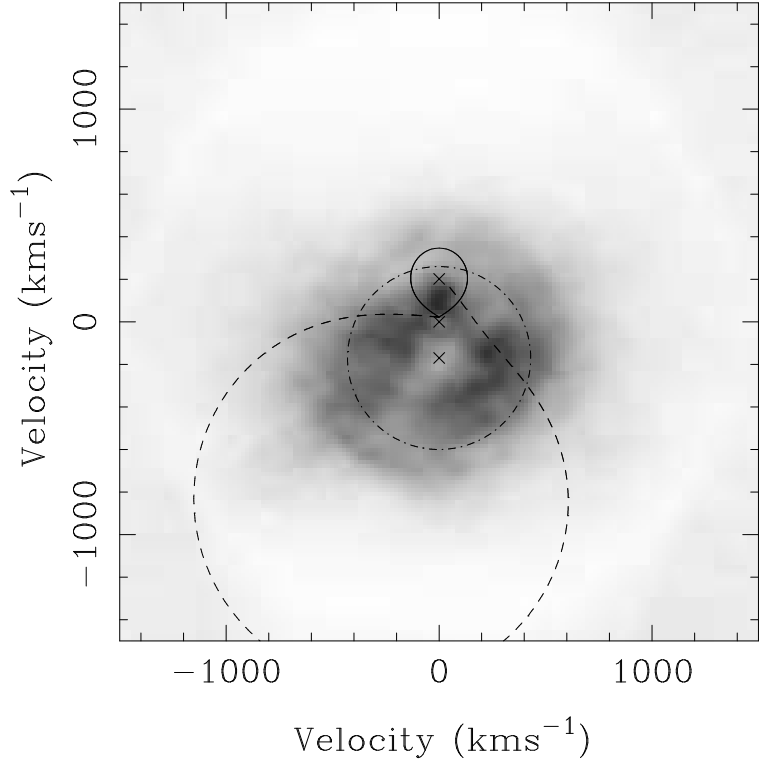}
\hspace*{\fill}
\caption{Doppler image of the H$\alpha$ line. Also plotted
 are the centre of mass of the donor star at V$_{y}=202\kms$ (upper
cross), the centre of mass of the system (middle cross), and the
centre of mass of the white dwarf (lower cross). The Roche lobe of the
mass donor star is plotted as well.  The dashed line shown is the
velocity path of the gas stream. The dash-dotted circle centred on
(0,-K$_{1}$) is the Keplerian velocity (430$\kms$) at the outer edge
of a disc of outer radius 0.8R$_{L_{1}}$.}
\label{fig:dopp}
\end{figure*}

Doppler tomography is an imaging technique which allows the
distribution of line emission to be mapped in velocity space (Marsh \&
Horne, 1988). A thorough analysis of several applications of this
method to real CV data are given in Marsh \& Horne (1990) and Marsh et
al.\,(1990).

Fig.~\ref{fig:dopp} shows the reconstructed Doppler tomogram of the
H$\alpha$ emission line in EM~Cyg. Any spectra affected by eclipse
were not included in the fit. The Doppler image shows a diffuse ring-like
emission structure, representing the accretion disc, which if tracking
the motion of the white dwarf should be centred at position (0,
-K$_{1}$) in velocity space. The inner edge of this ring represents
the velocity at the outer edge of the disc, which appears to be at a
velocity of $\approx$200$\kms$, which is low compared with values of
the outer disc velocities of other dwarf novae (e.g. Marsh \& Horne,
1990); the dashed circle drawn at 430$\kms$, is the Keplerian velocity
at the outer edge of a disc of radius 0.8R$_{L_{1}}$ (see
section~\ref{sec-mass}).  The apparent `hole' near the centre-of-mass
velocity of the system is due to the third star. This was not
removed from the original spectra in order to show its relative
magnitude and to demonstrate that we can pick up its H$\alpha$ flux.

\begin{figure*}
\hspace*{\fill}
\psfig{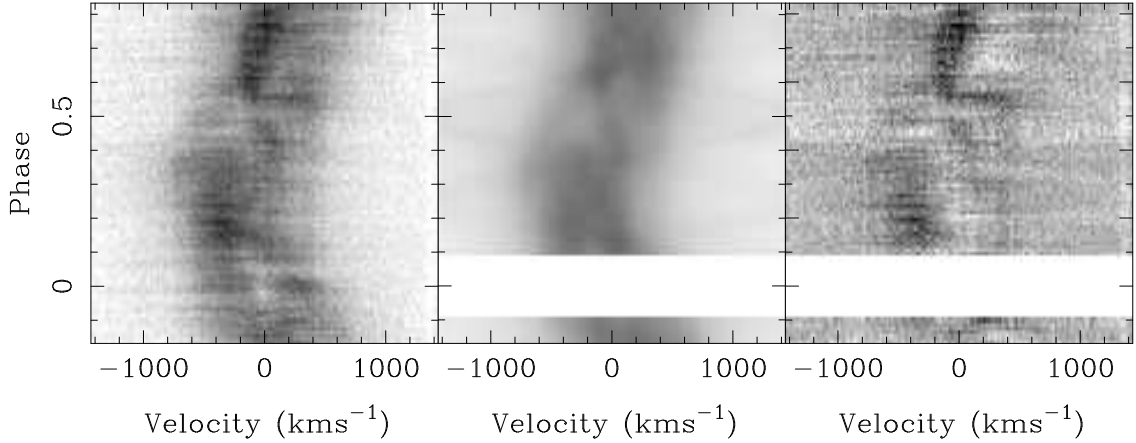}
\hspace*{\fill}
\caption{Trailed spectra of the H$\alpha$
emission line. Emission
sites are black on this plot. The left-hand panel shows the observed
data  used to calculate the Doppler tomogram. The
middle panel shows the trailed spectrum reconstructed from the
tomogram. Finally, the right-hand panel displays the residual image
formed when the computed data is subtracted from the observed data.}
\label{fig:hatrails}
\end{figure*}

\begin{figure*}
\hspace*{\fill}
\psfig{file=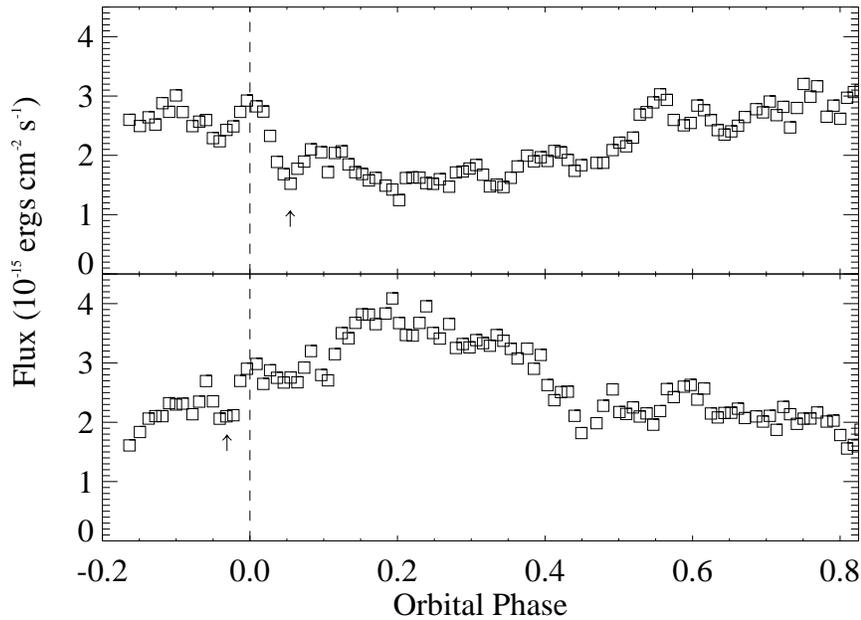,width=120mm}
\hspace*{\fill}
\caption{Variation of flux from the H$\alpha$ emission line with
orbital phase; 1) The upper plot shows the flux coming from the region
between 150 and 600 \protect$\kms$ 2) The lower plot shows the flux
coming from the velocity region -150 -- -600 $\kms$. The arrows indicate
the eclipses of the approaching (lower plot) and receding (upper plot)
limbs of a rotating disc.}
\label{fig:haligs}
\end{figure*}

The trailed spectra in the left-hand panel of Fig.~\ref{fig:hatrails}
shows the set of spectra used to compute the
Doppler image. Visible in Fig.~\ref{fig:hatrails} at phase 0 (with the
fiducial phase being defined as the inferior spectroscopic conjunction
of the mass donor) is the phenomenon of blue-shifted emission
being eclipsed before the red-shifted emission (Fig.~\ref{fig:haligs}),
known to be caused by a rotating disc; we use this in the next section
to constrain the orbital inclination. Although there is no obvious
S-wave from the mass donor visible in the trail, there is
emission on the irradiated face of the donor visible on the Doppler
map (Fig.~\ref{fig:dopp}) . The central panel of
Fig.~\ref{fig:hatrails} shows the  data computed from the 
H$\alpha$  Doppler map. The large region of emission around
phase 0.7 apparent in the actual data, is not completely
reconstructed in the computed data. Finally, the
right-hand panel shows the residual formed when the computed data is
subtracted from the observed data.

\section{Discussion}

\subsection{New System Parameters}

The main result of this study has been to discover that the spectrum
of EM~Cyg is contaminated by light from a star with a very similar
spectral type to its own mass donor. Removal of the contamination by
this star from the spectral lines in the region blueward of H$\alpha$
increases the measured semi-amplitude of the radial velocity curve of
the mass-losing star by over 50$\kms$, effectively removing the
problem previously associated with EM~Cyg: that the system should be
in a state of dynamically unstable mass
transfer. Table~\ref{tab:syspar} shows the re-measured orbital
elements, and those parameters which can be derived from them.

\begin{table*}
\begin{minipage}{120mm}
\begin{center}
\caption{New orbital and derived system parameters for EM Cygni}
\label{tab:syspar} 
\begin{tabular}{lcc}
\hline
\hline
\multicolumn{3}{c}{Orbital Elements\footnote{e = 0  (assumed)}} \\
\hline
Name & Value & Reference \\
\hline
Period (days) & 0.290909$\pm$0.0000042 & Robinson (1974) \\
K$_{2}$ & 202$\pm3\kms$ & this paper \\
K$_{1}$ & 170$\pm10\kms$ & Stover et al. (1981) \\
T$_{0}$ (JD$_{\odot}$)(2450621+) & 0.483$\pm$0.001 & this paper \\
$\gamma$\,($\kms$) & -25$\pm$2$\kms$ & this paper \\
$a$\,sin$i$ (cm) & $(1.52\pm0.04)\times 10^{11}$  & this paper \\
\hline
\multicolumn{3}{c}{Derived System Parameters} \\
\hline
$\vsini$ ($\kms$) & 140$\pm$6 & this paper \\
q (=M$_{2}$/M$_{1}$) & 0.88$\pm$0.05 & this paper \\
M$_{2}\,{\rm sin^{3}}i$ (M$_{\odot}$) & 0.77$\pm$0.08 & this paper \\
M$_{1}\,{\rm sin^{3}}i$ (M$_{\odot}$) & 0.88$\pm$0.05 & this paper \\
inclination, $i$ (degrees) & 61$\degr \leq i \leq 69\degr$ & this
paper \\
M$_{2}$ (M$_{\odot}$) & 0.99$\pm$0.12 & this paper \\
M$_{1}$ (M$_{\odot}$) & 1.13$\pm$0.08 & this paper \\
R$_{2}$ (R$_{\odot}$) & 0.87$\pm$0.07 & this paper \\ 
\hline
\end{tabular}
\vspace{-7mm}
\end{center}
\end{minipage}
\end{table*}

The range for the inclination $i$, shown in Table~\ref{tab:syspar} was
deduced as follows. A simulated H$\alpha$ trailed spectrum was created
using the system parameters determined previously (namely q and
$K_{1}+K_{2}$ as 0.88 and 372$\kms$ respectively) over the phase range
covered by the actual data, i.e. phases $-0.165\ {\rm to}\ 0.825$. The
power law used for the intensity over the disc as a function of radius
was -1.5, and the velocity limits of the plots were $\pm1500\kms$. The
outer disc radius used was 0.8R$_{L_{1}}$.  The eclipse of the disc in
the simulated H$\alpha$ trail becomes visible at inclination angle, $i
= 61\degr$. The inclination angle at which the eclipse matches that
seen in the actual data is $i = 67\degr$. At $i = 70\degr$, the white
dwarf is eclipsed, but it is apparent from our data that we are not
observing an eclipse of the white dwarf (Robinson, 1974). This places
an upper limit to the inclination of 70\degr. No part of the primary
Roche lobe is occulted for inclinations less than 61\degr, thus
placing a firm lower limit on $i$, higher than that reached by
Robinson (1974). Henceforth, we adopt a value of $i=67\degr\pm2\degr$
to cover the range where the uncertainty should be interpreted as a
1$\sigma$ estimate.

The radial velocity semi-amplitude of the white dwarf star was checked
using the double Gaussian technique developed by Schneider \& Young
(1980). The line wings are used to determine the radial velocities
because they form in an area which is close to the white dwarf and so
disruption of symmetry by the gas stream and mass donor star are
minimised. Twin Gaussians (200$\kms$ FWHM) with separations varying
from 1200 -- 1800 $\kms$ were used. This procedure gave consistent
values of the white dwarf radial velocity semi-amplitude $K_{1} = 171
\pm 4\kms$, in agreement with that obtained by Stover et al.\
(1981). The systemic velocity measured from the radial velocity curve
fits was $\gamma = -26\pm5\kms$, consistent with that obtained from
the mass donor analysis. The emission line radial-velocity
curves were measured to be delayed by $10\degr\pm3\degr$ from being
exactly in anti-phase with with the donor star.

\subsection{Masses}
\label{sec-mass}

The measurements of $\vsini$ and K$_{2}$ can be used to
deduce the mass ratio $q=M_{2}/M_{1}$. Assuming that the donor star
rotates synchronously then 

\begin{equation}
\frac{v\,{\rm sin}i}{K_{2}} = \frac{R_{2}(1+q)}{a}.
\label{eq:vsini}
\end{equation}

The relative size of the donor star is constrained  by Roche geometry
to be
\begin{equation}
\frac{R_{2}}{a} = \frac{0.49q^{2/3}}{0.6q^{2/3}+{\ln}(1+q^{1/3})},
\hspace*{0.5in} {\rm for}\;0< q < \infty,
\label{eq:eggleton}
\end{equation}
Eggleton (1983). Applying these equations we find
$q=0.88\pm0.05$. This is consistent with that found from K$_{1}$:
$q=K_{1}/K_{2}=0.84\pm0.06$, but we prefer the value determined from
$\vsini$ as K$_{1}$ measurements often suffer distortions
(Stover, 1981).
 
The values obtained here for K$_{2}$, q (derived from the rotational
broadening measurement) and $i$ can be combined with the
orbital period obtained by Robinson (1974) to calculate the
masses of each component using:

\begin{equation}
M_{1} = \frac{P_{orb}K_{2}^{3}(1+q)^{2}}{2\pi G{\rm sin^{3}}i}
\end{equation}
\begin{equation}
M_{2} = \frac{P_{orb}K_{2}^{3}(1+q)^{2}q}{2\pi G{\rm sin^{3}}i}.
\end{equation}

The values obtained are shown in Table~\ref{tab:syspar}.  Assuming the
inclination value concluded in the previous section, $i=67\degr\pm2\degr$,
we calculate component masses of $M_{1}=1.12\pm0.08M_{\odot}$ and
$M_{2}=0.99\pm0.12M_{\odot}$. The value for the white dwarf mass is
consistent with those given for other Z Cam systems (Ritter \& Kolb,
1998). However, the mass for the donor star is close to that of the
Sun, which is a G2V star, compared with our observed K3 (Allen
1973). Beuermann et al.\ (1998) discuss whether the mass donor stars
in CVs are main-sequence stars, using relationships between spectral
type and orbital period in a sample of CVs. They show that the donor
stars in CVs with shorter orbital periods fit the `unevolved'
evolutionary tracks well, whilst those at longer periods tend to fit
`evolved' models better. EM~Cyg with its orbital period of 6.98 hours,
and mass donor spectral type of K3, fits on the plot nearer to the
`moderately evolved main-sequence' track, indicating that EM~Cyg may
have an evolved mass donor, because it appears to be oversized and
over-massive for its spectral type. The radius of the mass donor is
$R_{2}=0.87R_{\odot}\pm0.07$ which compares with 0.74R$_{\odot}$ for a
K5V star, and 0.85R$_{\odot}$ for a K0V star (Allen, 1973), indicating
that it is oversized compared to a main-sequence star of the same
spectral type. However, the mass deduced for the donor star
($0.99\pm0.12M_{\odot}$) is not consistent with its spectral type
according to the current theoretical models of Kolb \& Baraffe (1999). Their
stellar models only become as cool as K3 for a donor star mass of
$\lse 0.85M_{\odot}$, and this is only if the mass loss rate has been
rather high (and the mass donor is unevolved). In addition, they show
that if the mass donor is `evolved' it would have a slightly earlier
spectral type than a ZAMS star with a mass of 1 M$_{\odot}$. In
summary, the new system parameters appear to be inconsistent with the
current CV mass-donor models.

We have identified a third star contaminating the spectral region
around H$\alpha$ wavelengths in EM~Cyg, causing the absorption lines
in this area to appear narrow, with none of the broadening present
which is associated with the actual mass donor. The spectral type of
this third star has been identified as K2V -- K5V, which is similar to
that of the mass donor. It also contributes a similar percentage
of light to the spectrum of EM~Cyg as the actual mass donor. The
third star contributes a fraction 0.160$\pm$0.002 to the light
of EM~Cyg at H$\alpha$ wavelengths, whilst that contributed by the
mass donor is 0.231$\pm$0.005.  Once this third star has been accounted
for, the measurements for the radial velocity semi-amplitude increase
(from $135\pm3\kms$ to $202\pm3\kms$) and the mass ratio of EM~Cyg falls
below one, removing the conflict with the theory. The semi-amplitude
of the radial velocity curve for the white dwarf is $171\pm4\kms$,
which agrees with the value $170\pm10\kms$ given by Stover et al.\
(1981), and which gives a mass ratio of q=0.88$\pm$0.05$\kms$. The
radial velocity of the centre-of-mass of the binary from the
absorption line measurement is $-25\pm2\kms$. The radial velocity of
the contaminating spectrum is also consistent with this value
($-20\pm3\kms$) which implies that the third star and EM~Cyg may be physically
associated. In addition, the similarity of spectral type and
fractional contribution to the spectrum of EM~Cyg indicates that the
third star and the CV may be at similar distances. However, the field
of EM~Cyg is quite a crowded one, and so a chance superposition is also 
plausible. To address this we acquired some direct images on the
William Herschel Telescope which have a FWHM of 0.7 arcsec, but we
still could not resolve the third star. 
The value of $a\,{\rm sin}i$ obtained with the new semi-amplitude and
mass-ratio values is $1.52\times10^{11}$cm.  Assuming Keplerian motion
in the disc, then
\begin{equation}
\frac{r_{\rm d}}{R_{L_{1}}} =\frac{a}{R_{L_{1}}} \frac{GM_{1}{\rm
sin}^{3}i}{a{\rm sin}i\,(v_{\rm disc}\,{\rm sin}\,i)^{2}},
\label{eq:disc}
\end{equation}
where R$_{L_{1}}$ is the distance from the centre of mass of the
white dwarf to the inner Lagrangian point and is equal to $0.51a$ for
$q=0.88$. 
Using the determined values of M$_{1}{\rm
sin}^{3}i$=0.88$\pm$0.05M$_{\odot}$, $a\,{\rm sin}\,i$ =
$(1.52\pm0.04)\times10^{11}$ cm, and the outer disc velocity, $v_{\rm
disc}{\rm sin}i$ as 305$\pm30\kms$ we calculate a value of r$_{{\rm
d}}/R_{L_{1}}=1.61$. This is physically impossible as it implies that
the disc must occupy space outside the Roche lobe of the white
dwarf. For
a more believable disc radius of 0.8R$_{L_{1}}$, the predicted outer
disc velocity, using equation~\ref{eq:disc} is 430$\kms$. Stover
et al.\ (1981) measured the half-separation of the peaks of the
H$\beta$ line to be $390\pm15\kms$, giving an outer disc radius of
0.98R$_{L_{1}}$, still too large, but not as discrepant as H$\alpha$,
for which we have no explanation.

\section{Conclusions}
We have found a significant contribution to the spectrum of EM~Cyg in
the H$\alpha$ wavelength region from a K-type star other than the mass
donor. When removed from the spectrum, the absorption line radial
velocity amplitude increases by over 60 $\kms$ to $202\pm3\kms$
bringing the mass ratio calculated from the radial velocity
semi-amplitudes down to 0.88 from the value of 1.26 calculated by
Stover et al.\ (1981). This solves the `mystery' surrounding the
supposed dynamical instability EM~Cyg would suffer if the mass of the
donor star was less than 0.8M$_{\odot}$ and more massive than the
white dwarf (as deduced by Stover et al., 1981).  The radial velocity
of this `contaminating' third star ($-20\pm3\kms$) is similar to the
measured radial velocity of the centre-of-gravity of EM~Cyg
($-25\pm2\kms$), which is consistent with the stars being physically
associated. In addition, the spectral types of the two red stars are
similar (K2-K5V). The third star contributes 16.0$\pm$0.2 per cent
of the light in this wavelength region, 7 per cent less than that
contributed by the donor star (23.1$\pm$0.5 per cent). The rotational
broadening of the mass donor in EM~Cyg is measured to be
$140\pm6\kms$, and its mass calculated as
$0.99\pm0.12M_{\odot}$. Smith \& Dhillon (1998) plot the mass (in
solar mass units) versus the spectral type of the mass donor in a
sample of CVs. EM~Cyg is on this plot with a spectral type of K5V, and
a mass of $0.76\pm0.10M_{\odot}$. The new value of M$_{2}$ puts the
system above where it should be were it a main-sequence star,
confirming that the mass donor in EM~Cyg must be evolved.  If the
third star is actually physically associated with EM~Cyg, then
this would make it only the second CV known to be in a triple system
(Reimers, Griffin \& Brown 1988).

\section*{Acknowledgments}
TRM was supported by a PPARC Advanced Fellowship during the course of
this work.  The data reduction and analysis were carried out on the
Southampton node of the UK STARLINK computer network. The INT and WHT
are operated on the island of La Palma by the Isaac Newton Group in
the Spanish Observatorio del Roque de los Muchachos of the Instituto
de Astrofisica de Canarias. In this research, we have used, and
acknowledge with thanks, data from the AAVSO International Database,
based on observations submitted to the AAVSO by variable star
observers worldwide.

\end{document}